%% file: ms.tex
\documentclass[iop]{emulateapj}
\usepackage[citebordercolor={0 .5 .5}]{hyperref}
\usepackage{multirow}

\newcommand{\beq}{\begin{equation}}
\newcommand{\eeq}{\end{equation}}
\newcommand{\be}{\begin{equation}}
\newcommand{\ee}{\end{equation}}
\newcommand{\bea}{\begin{eqnarray}}
\newcommand{\eea}{\end{eqnarray}}
\newcommand{\bdi}{\begin{equation}}
\newcommand{\edi}{\end{equation}}

\newcommand{\HI}{H\,{\sc i}} 
\newcommand{\HII}{H$_2$}

\newcommand{\rmicron}{$\,\micro$m}

\def\lsim{\,\lower2truept\hbox{${<\atop\hbox{\raise4truept\hbox{$\sim$}}}$}\,}
\def\gsim{\,\lower2truept\hbox{${>\atop\hbox{\raise4truept\hbox{$\sim$}}}$}\,}

\usepackage{amssymb,amsmath}
\usepackage{natbib}  
\usepackage[mediumspace,Gray,squaren]{SIunits} 

\shorttitle{The Herschel Stripe 82 Survey (HerS)}
\shortauthors{Viero et al.}

\begin{document}
\title{The Herschel Stripe 82 Survey (HerS): Maps and Early Catalog$^{\dagger}$}
\author{M.P.~Viero\altaffilmark{1,$\ddagger$},
V.~Asboth\altaffilmark{2},
I.~G.~Roseboom\altaffilmark{3},
L.~Moncelsi\altaffilmark{1},
G.~Marsden\altaffilmark{2},
E.~Mentuch Cooper\altaffilmark{4},
M.~Zemcov\altaffilmark{1,5},
G.~Addison\altaffilmark{2},
A.J.~Baker\altaffilmark{6},
A.~Beelen\altaffilmark{7},
J.~Bock\altaffilmark{1,5},
C.~Bridge\altaffilmark{1},
A.~Conley\altaffilmark{8},
M.~J.~Devlin\altaffilmark{9},
O.~Dor\'e\altaffilmark{5,1},
D.~Farrah\altaffilmark{10},
S.~Finkelstein\altaffilmark{4},
A.~Font-Ribera\altaffilmark{11,12},
J.~E.~Geach\altaffilmark{13},
K.~Gebhardt\altaffilmark{4},
A.~Gill\altaffilmark{8},
J.~Glenn\altaffilmark{14,8},
A.~Hajian\altaffilmark{15},
M.~Halpern\altaffilmark{2},
S.~Jogee\altaffilmark{4},
P.~Kurczynski\altaffilmark{6},
A.~Lapi\altaffilmark{16,17},
M.~Negrello\altaffilmark{18},
S.~J.~Oliver\altaffilmark{19},
C.~Papovich\altaffilmark{20},
R.~Quadri\altaffilmark{21,22},
N.~Ross\altaffilmark{12},
D.~Scott\altaffilmark{2},
B.~Schulz\altaffilmark{1,23},
R.~Somerville\altaffilmark{6},
D.~N.~Spergel\altaffilmark{24},
J.~D.~Vieira\altaffilmark{1},
L.~Wang\altaffilmark{25},
R.~Wechsler\altaffilmark{26}}
\altaffiltext{$\ddagger$}{Email: marco.viero@caltech.edu}
\altaffiltext{$\dagger$}{Herschel is an ESA space observatory with science instruments provided by European-led Principal Investigator consortia and with important participation from NASA.}
\altaffiltext{1}{California Institute of Technology, 1200 E. California Blvd., Pasadena, CA 91125}
\altaffiltext{2}{Department of Physics \& Astronomy, University of British Columbia, 6224 Agricultural Road, Vancouver, BC V6T~1Z1, Canada}
\altaffiltext{3}{Institute for Astronomy, University of Edinburgh, Royal Observatory, Blackford Hill, Edinburgh EH9 3HJ, UK}
\altaffiltext{4}{Department of Astronomy, The University of Texas at Austin, Austin, TX 78712}
\altaffiltext{5}{Jet Propulsion Laboratory, 4800 Oak Grove Drive, Pasadena, CA 91109}
\altaffiltext{6}{Department of Physics and Astronomy, Rutgers, The State University of New Jersey, 136 Frelinghuysen Rd, Piscataway, NJ 08854}
\altaffiltext{7}{Institut d'Astrophysique Spatiale (IAS), b\^atiment 121, Universit\'e Paris-Sud 11 and CNRS (UMR 8617), 91405 Orsay, France}
\altaffiltext{8}{Center for Astrophysics and Space Astronomy 389-UCB, University of Colorado, Boulder, CO 80309}
\altaffiltext{9}{Department of Physics and Astronomy, University of Pennsylvania, Philadelphia, PA 19104}
\altaffiltext{10}{Department of Physics, Virginia Tech, Blacksburg, VA 24061}
\altaffiltext{11}{Institute of Theoretical Physics, University of Zurich, Winterthurerstrasse 190, 8057 Zurich, Switzerland}
\altaffiltext{12}{Lawrence Berkeley National Laboratory, 1 Cyclotron Road, Berkeley, CA 94720}
\altaffiltext{13}{Centre for Astrophysics Research, Science \& Technology Research Institute, University of Hertfordshire, Hatfield, AL10 9AB, UK}
\altaffiltext{14}{Dept. of Astrophysical and Planetary Sciences, CASA 389-UCB, University of Colorado, Boulder, CO 80309}
\altaffiltext{15}{Canadian Institute for Theoretical Astrophysics, University of Toronto, Toronto, ON\ M5S~3H8, Canada}
\altaffiltext{16}{Dip. Fisica, Univ. \lq Tor Vergata\rq, Via Ricerca Scientifica 1, 00133 Roma, Italy}
\altaffiltext{17}{Astrophysics Sector, SISSA, Via Bonomea 265, 34136 Trieste, Italy}
\altaffiltext{18}{INAF, Osservatorio Astronomico di Padova, Vicolo Osservatorio 5, I-35122 Padova, Italy}
\altaffiltext{19}{Astronomy Centre, Dept. of Physics \& Astronomy, University of Sussex, Brighton BN1 9QH, UK}
\altaffiltext{20}{George P. and Cynthia Woods Mitchell Institute for Fundamental Physics and Astronomy, Department of Physics and Astronomy, Texas A\&M University, College Station, TX 77843}
\altaffiltext{21}{Hubble Fellow}
\altaffiltext{22}{Carnegie Observatories, Pasadena, CA 91101}
\altaffiltext{23}{Infrared Processing and Analysis Center, MS 100-22, California Institute of Technology, JPL, Pasadena, CA 91125}
\altaffiltext{24}{Joseph Henry Laboratories of Physics, Jadwin Hall, Princeton University, Princeton, NJ\ 08544}
\altaffiltext{25}{Institute for Computational Cosmology, Department of Physics, University of Durham, South Road, Durham, DH1 3LE, UK}
\altaffiltext{26}{Kavli Institute for Particle Astrophysics and Cosmology and Department of Physics, Stanford University, 382 Via Pueblo Mall, Stanford, CA 94305}
\begin{abstract}
We present the first set of maps 
and band-merged catalog from the \emph{Herschel} Stripe 82 Survey (HerS).  
Observations at 250, 350, and 500\rmicron\ were taken with the Spectral and Photometric Imaging Receiver (SPIRE) instrument aboard the \emph{Herschel Space Observatory}.  %
HerS covers $79\, \rm deg^2$  along the SDSS Stripe 82 to an average depth of 13.0, 12.9, and $14.8\, \rm mJy\, beam^{-1}$ (including confusion) at 250, 350, and 500\rmicron, respectively.     
HerS was designed to measure correlations with external tracers of the dark matter density field --- either point-like (i.e., galaxies selected from radio to  X-ray) or extended (i.e., clusters and gravitational lensing) --- in order to measure the bias and redshift distribution of intensities of infrared-emitting dusty star-forming galaxies and AGN.    
By locating HeRS in Stripe 82, we maximize the overlap with available and upcoming cosmological surveys. 
The band-merged catalog contains $3.3\times 10^4$  sources detected at a significance of $\gsim 3\sigma$ (including confusion noise). 
The maps and catalog are available at  \url{http://www.astro.caltech.edu/hers/}.
\end{abstract}
 

\keywords{cosmology: observations, submillimeter: galaxies -- infrared: galaxies -- galaxies: evolution -- large-scale structure of universe}

\section{Introduction}
\label{sec:intro}
\begin{figure*}[t!]
\centering
\hspace{10mm}
\includegraphics[width=1.0\textwidth]{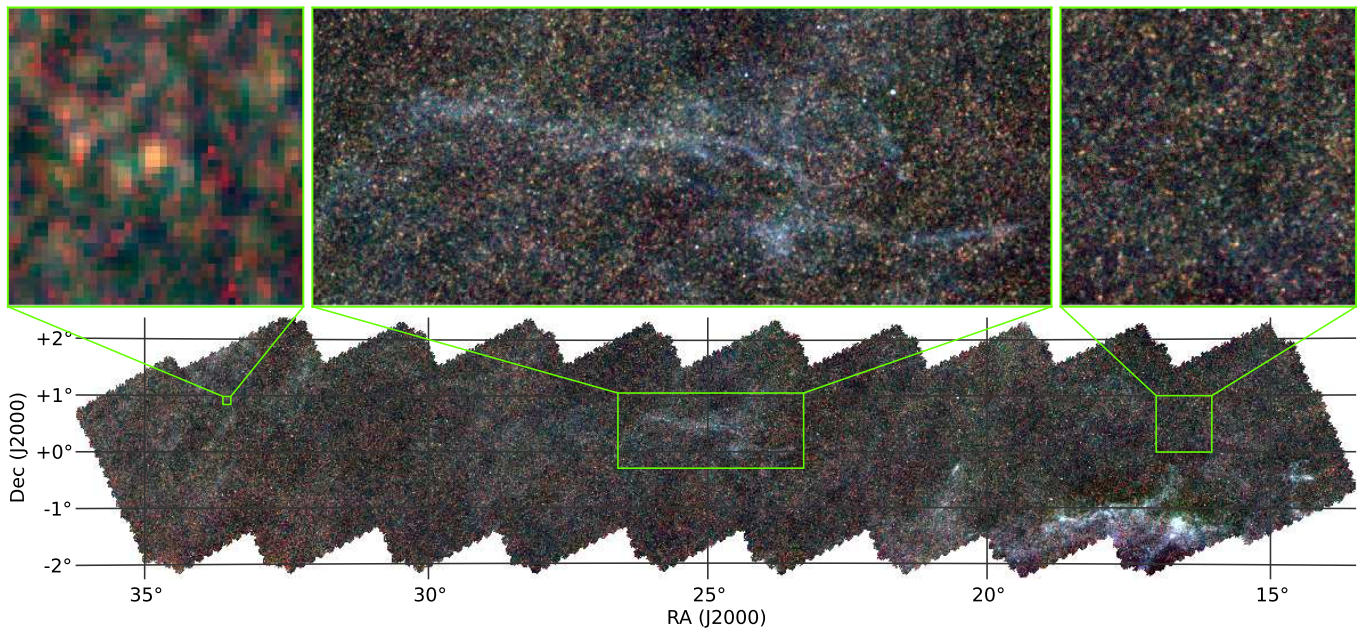}
\caption{Three-color image of the HerS field with 250, 350, and 500\rmicron\ as blue, green, and red, respectively.  Note that 250 and 350\rmicron\ maps were convolved so that all three maps have the same angular resolution.  
{\bf Left Panel:} A high-redshift candidate \lq\lq red peaker\rq\rq, with $S_{250} < S_{350} < S_{500}$, such that its SED suggests it lies somewhere between \emph{z} of 3 and 7.
{\bf Center Panel:} A foreground cloud of Galactic cirrus (see \S~\ref{fig:cirrus}), with column densities reaching $N_{\rm H} \sim 4.5 \times 10^{21}\, \rm cm^{-2}$. 
{\bf Right Panel:} A typical $1^{\circ}\times 1^{\circ}$ \lq\lq blank field\rq\rq, which contains mostly dusty star-forming galaxies at intermediate to high redshifts.  
}
\label{fig:hers}
\end{figure*}
The cosmic infrared background (CIB) traces the star-formation history of the Universe; roughly half the emission of young stars appears in the ultraviolet and optical, while the rest is absorbed by dust and then emitted at 
far-infrared wavelengths 
 \citep[][]{puget1996,fixsen1998,hauser2001,dole2006}. 
 Over the last decade a key goal of far-IR/submillimeter astronomy has been to identify the galaxies that produce the CIB.  
 Recent deep surveys with the Balloon-borne Large Aperture Submillimeter Telescope \citep[BLAST;][]{devlin2009,marsden2009,pascale2009} 
and the \emph{Herschel Space Observatory} \citep[H-ATLAS, HerMES, PEP;][]{eales2010, oliver2012, lutz2011}
 as well as ground-based submillimeter facilities such as LABOCA 
  \citep[LESS;][]{weiss2009} and SCUBA-2 \citep{geach2013}  
 have ``resolved'' over 80\% of the CIB at submillimeter wavelengths, via direct counting of sources 
 \citep{oliver2010,geach2013}, \emph{P(D)} techniques \citep{glenn2010}, and stacking \citep{dole2006,berta2011, bethermin2012b,  viero2013a}. 
The resolution of this large fraction of the CIB into individual sources makes it clear that the CIB, at least near to its peak at $\sim 200$\rmicron, is dominated by a moderate luminosity population  
 \citep[i.e., $L_{\rm IR}\leq10^{12}\,$L$_{\odot}$;][]{bethermin2011,wang2013}  
 in the broad redshift interval $1\leq z\leq3$ \citep[e.g.,][]{viero2013b}. 
Additionally, measurements of the CIB power spectrum \citep[e.g.,][]{amblard2011, lagache2011,lagache2013, viero2009,viero2013a}
yield estimates of the source clustering properties.  
 
While the determination of these broad characteristics represents a remarkable achievement, much remains to be done to link the CIB, and the infrared (IR) luminous galaxies which make it up, to the general galaxy population. This goal requires determining the multi-wavelength characteristics of galaxies detected at far-IR/submillimeter wavelengths, and hence the physical properties that these wavelengths probe, e.g., rest-frame optical light tracing stellar mass, X-ray tracing black hole accretion, etc. A major complication is that the confusion-limited sensitivity of single-dish far-IR/submillimeter facilities is such that only the most luminous sources (i.e., $L_{\rm IR}\geq10^{12}\,$L$_{\odot}$) can be individually detected in the key redshift range $1\leq z\leq3$. Interferometric facilities like ALMA are not limited in this way, although their small fields of view (e.g. $\ll 1$ arcmin$^2$) means that large blind surveys of the IR-galaxy population are inefficient and prohibitively expensive. 
To characterize the physical properties of the galaxies that dominate the CIB will instead require the use of statistical techniques, i.e., stacking or similar \citep[][]{devlin2009,marsden2009,pascale2009,kurczynski2012,viero2012,viero2013b,roseboom2012}, and hence very large numbers ($>100{,}000$) of galaxies detected at wavelengths with higher resolution (typically optical/near-IR).

Motivated by the importance of the CIB and the need to have large multi-wavelength surveys to understand its properties, we have conducted  the {\it Herschel} Stripe 82 Survey (HerS; Figure~\ref{fig:hers}). 
HerS consists of $79\, \rm deg^2$ of contiguous imaging with the SPIRE instrument \citep{griffin2010} on the \emph{Herschel Space Observatory} \citep{pilbratt2010}  to roughly the confusion limit \citep[$\sim 7\, \rm mJy$ at the wavelengths 250, 350, and 500\rmicron;][]{nguyen2010}.   
Crucially, HerS is positioned to overlap with a rich array of both existing and planned 
galaxy surveys 
 in the Sloan Digital Sky Survey's \citep[SDSS;][]{york2000}  ``Stripe 82'' field, including: 
The SDSS-III's Baryon Oscillation Spectroscopic Survey \citep[BOSS;][]{eisenstein2011}, 
VICS82 (VISTA+CFHT Stripe 82 survey; Geach et al.\@ in prep.), 
VISTA-VIKING \citep{emerson2004}, 
VLA-Stripe82 \citep{hodge2011},  
The Hobby-Eberly Telescope Dark Energy Experiment \citep[HETDEX;][]{hill2008},  
The \emph{Spitzer}-HETDEX Exploratory Large Area Survey  \citep[SHELA;][]{papovich2012}, 
The \emph{Spitzer}-IRAC Equatorial Survey \citep[SpIES;][]{richards2012}, and  
Hyper Suprime-Cam  \citep[HSC;][]{miyazaki2012} surveys. 
The combination of SHELA/SpIES, which are \emph{Spitzer}-warm IRAC surveys of Stripe 82, and HETDEX, a wide-area spectroscopic survey targeting emission lines at $z>2$, will detect hundreds of thousands of galaxies and provide the key information required to interpret the HerS images.

In addition, HerS overlaps with a survey of the cosmic microwave background (CMB) conducted by The Atacama Cosmology Telescope \citep[ACT;][]{sievers2013} in  Stripe 82.  
The power in CMB maps on angular scales  $\ell \gsim 2000$ is dominated by point sources --- both dusty and/or radio \citep[e.g.,][]{vieira2010,das2011,reichardt2012} --- which act as foregrounds when attempting to study, for example, the damping tail of the CMB power spectrum \citep[e.g.,][]{keisler2011}, or the thermal and kinetic Sunyaev Zel'dovich (SZ) effect from clusters or reionization, respectively \citep[e.g.,][]{mcquinn2005}.   
Cross-correlations between the CIB and CMB provide critical constraints for models of this contamination \citep[e.g.,][]{hajian2012,lagache2013}, particularly on the smallest angular scales where the power spectrum is dominated by the non-linear, 1-halo term \citep[e.g.,][]{viero2013a}.    
As we will later address, determining this component is one of the main motivations for locating the survey in the Stripe, and thus drives some of the mapmaking decisions.  

\begin{figure*}[ht!]
\centering
\hspace{10mm}
\includegraphics[width=1.0\textwidth]{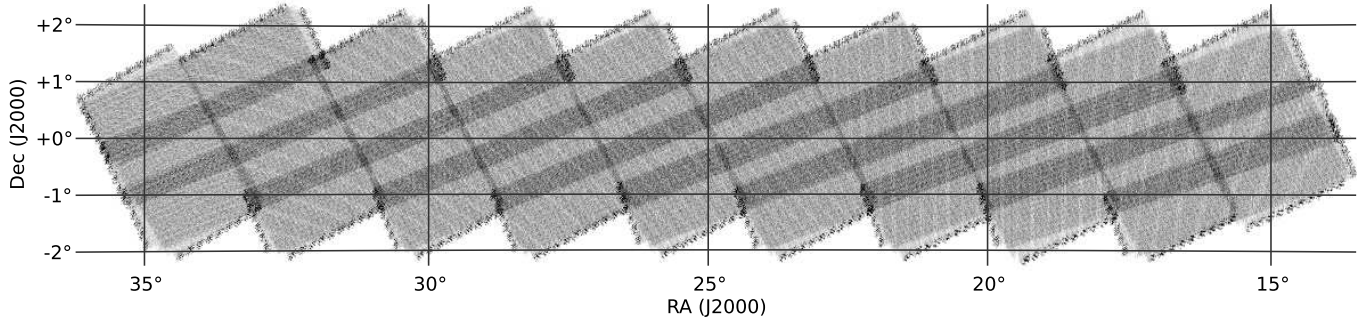}
\caption{Coverage map of the 350\rmicron\ data.  
The majority of the map is covered twice, while the dark grey stripes are the regions covered three times.   
As the scan orientations of the telescope at the ecliptic are fixed irrespective of the observing season,  this scan strategy was chosen to guarantee complete coverage of the area along the Stripe. }
\label{fig:hits}
\end{figure*}
In addition to these large statistical analyses, the large area of HerS adds an additional 79\,deg$^2$ to the existing wide-area H-ATLAS and HerMES surveys 
to identify and study sources that are ``rare'' on the sky. The HerS field contains tens of nearby luminous IR galaxies (LIRGs; $L_{\rm IR}\geq10^{11}\,$L$_{\odot}$) that are close enough to be resolved by {\it Herschel} at 250\rmicron. Meanwhile, we expect to identify close to 100 distant ($z>2$) galaxies with very high observed luminosities, with many of these resulting from lensing by foreground galaxies \citep[like those found in e.g.][]{negrello2010,wardlow2013,vieira2013}. 
Finally, HerS will contain many thousands of LIRGs at intermediate redshifts, making it a rich dataset for the study of IR-luminous galaxy evolution since $z=1$.

This paper describes the first release of HerS maps and catalog, including design strategy (\S~\ref{sec:design}), mapmaking and map properties (\S~\ref{sec:maps}), and catalog construction and statistics (\S~\ref{sec:catalog}).  Data are available at \url{ http://www.astro.caltech.edu/hers/}.

\section{Survey Design}
\label{sec:design}
HerS was  designed  to optimize cross-correlation measurements 
with ancillary data sets.   
This objective requires two key ingredients: well understood ancillary data (preferably of high source density); and submillimeter maps covering large areas with faithful reconstruction of large scales.      
To satisfy the first criterion, the survey was located in Stripe 82 which, in addition to the numerous surveys already described, 
will uniquely be observed by both HETDEX and ACT.  Furthermore, its equatorial location --- visible from most ground-based telescopes --- makes it well-placed to be a valuable legacy field in the future.  
Its location was driven by the relatively low Galactic cirrus foreground (e.g., $N_{\rm H}\sim 1.7\times 10^{21}\, \rm cm^{-2}$; see \S~\ref{sec:cirrus}) with respect to the rest of the Stripe.  Combined with the HeLMS survey \citep[the largest field in HerMES;][]{oliver2012}, the full $\sim 150\, \rm deg^2$ of Stripe 82 with $N_{\rm H} \le  3\times 10^{21}\, \rm cm^{-2}$ has been imaged.  

The second criterion --- the need for large areas --- is again due to source confusion.  As shown in  e.g., \citet{acquaviva2008}, the signal-to-noise ratio in cross-correlation measurements is proportional to the square root of $f_{\rm sky}$, or areal coverage, and is inversely proportional to the square root of the noise.  For the case of maps observed with SPIRE,  since the noise as a function of observing time quickly approaches the confusion limit, observation time is more optimally spent going wider rather than deeper.  
To reconstruct the largest scales, the maps were imaged in fast-scan mode ($60\, \rm arcsec\, s^{-1}$) and cross-linked with nearly orthogonal scans.  
The equatorial location of the field limited the orientations possible with the telescope.  Coverage of the Stripe, visible in the coverage map shown in Figure~\ref{fig:hits}, was achieved in 21 scans over 34.5 hours of observing time.  This scan pattern resulted in 10 stripes with additional coverage, i.e., 3 rather than 2 scans;  we address in later sections how these deeper stripes  affect the noise properties  of the maps and completeness properties of the catalogs.

\section{Maps}
\label{sec:maps}
Observations cover $79\, \rm deg^2$ in the equatorial Stripe 82, spanning  $13^{\circ}$ to $37^{\circ}$ ($\rm 0^{h}54^m$ to $\rm 2^{h}24^m$) in RA, and $-2^{\circ}$ to $ 2^{\circ}$ in declination.
Maps were made using the maximum likelihood mapmaker {\sc sanepic} \citep[Signal and Noise Estimation Procedure Including Correlations;][]{patanchon2008}. This mapmaker is optimized for datasets where a large number of detectors observe the same area of the sky and the correlated (or common-mode) noise between the time-ordered data (TOD, or timestream) of these detectors cannot be ignored. The main source of this common-mode noise is the drift in temperature of the cooler bath surrounding the detector arrays. Instead of removing all large-scale variations with high-pass filtering, as many other mapmakers do, {\sc sanepic} separates the low-frequency correlated noise from the sky signal, resulting in maps in which large-scale variations of the sky are better preserved. 

Two sets of maps at 250, 350, and 500\rmicron\ were made in order to accommodate different science goals. For the first set, we used a tangent plane (TAN) projection with pixel sizes of 6, 8.33 and 12\,arcsec for the 250, 350, and 500\rmicron\ maps, respectively. These values are typical for SPIRE maps, chosen to correspond to roughly one-third of the size of the SPIRE beams (18.1, 25.2 and 36.6\,arcsec full-width at half-maximum). 
 Since the HerS field overlaps with the equatorial Stripe observed by the Atacama Cosmology Telescope (ACT), we also made maps using the nominal ACT map projection for cross-analysis of the two data sets.  
 The motivation for matching pixels is that it avoids the reprojecting/regridding of maps that would be necessary to perform map-based operations --- whether in Fourier space or otherwise --- which could potentially  introduce systematic uncertainties.   
  The HerS-ACT maps were made using a cylindrical equal-area (CEA) projection with pixel sizes of 29.7\,arcsec in all three bands, corresponding to the nominal ACT pixel size.
\subsection{Data preprocessing}
\label{sec:dp}

The raw data from the bolometer arrays are stored as separate TODs for each detector. Before the data are fed into our mapmaker several preprocessing steps are applied to the raw TODs.  We used the HIPE \citep[Herschel Interactive Processing Environment;][]{ott2010}, version 11.0.1 mapmaking software package to convert the uncalibrated raw TODs into the so-called Level 1 format, which is the input format used by mapmakers. The preprocessing steps involve detecting jumps in the signal, flagging glitches, and correcting for the low-pass filter response of the electronics and for the bolometer time response. Calibration of the data also happens at this early processing stage. The Level 1 data are read in by the {\sc smap} mapmaking software package \citep{levenson2010, viero2013a} and exported to the format accepted by {\sc sanepic}. {\sc smap} also uses an additional iterative glitch detection algorithm during mapmaking, and the deglitching information can be re-used later. This existing deglitching-information from preliminary HerS maps created with {\sc smap} is also applied to our TODs. For details of these preprocessing steps see appendix A of \citet{viero2013a}. 
Both the HIPE and {\sc smap} pipelines have their own algorithms to remove temperature drifts on long timescales by fitting to thermistor TODs.
Since {\sc sanepic} is optimized to deal with large-scale correlated noise, we turn off the temperature drift removal step in HIPE and {\sc smap} during preprocessing.  The last preprocessing steps are applied by {\sc sanepic}. A first-order polynomial is fit to and removed from each data segment, because the variations on timescales longer than the timestream itself can cause leakage during Fourier-transformation, which would introduce artifacts in our maps. 
{\sc sanepic} fills any gaps in the TODs and the data segments are apodized at the edges over 50 samples. This measure is needed since the mapmaker assumes that the ends of each data segment are strongly correlated (``circular'').  
 
\subsection{Mapmaking}
\label{sec:mm}

The {\sc sanepic} mapmaking method is described in detail in \citet{patanchon2008}; here we review the salient points.
The timestream of a bolometer indexed by $\textit{i}$ can be modeled as
\begin{equation} d_i(t)= \sum\limits_p A_{ip}(t) s_p+n_i(t) ,
\end{equation}
where \textit{t} is the time when the sample was taken, $s_p$ is the signal in pixel \textit{p} of the map of the sky and $A_{ip}(t)$ is the pointing matrix, which gives the weight of the contribution of the signal in pixel $\textit{p}$ to the timestream of bolometer $\textit{i}$ at time $\textit{t}$. We assert that $s_p$ corresponds to the beam-convolved sky, in which case the pointing matrix tells us the position where bolometer $\textit{i}$ points on the sky at time  $\textit{t}$. The noise term $n_i(t)$, whose properties are assumed to be stationary, is the sum of two components: the uncorrelated noise between different detectors $\tilde{n}_i(t)$; and a common-mode signal, $\alpha_i c(t)$, seen by all detectors at a given time.  This \lq\lq noise\rq\rq\ term is 
\begin{equation}
n_i(t)=\tilde{n}_i(t)+ \alpha_i c(t),
\end{equation}
where \emph{c(t)} is the correlated noise which is the same for all detectors apart from a detector-dependent multiplicative factor $\alpha_i$.
The sky signal can be estimated from the detector TODs using maximum likelihood methods. The solution is given by 
\begin{equation}
\hat{s}=(A^{\rm T} N^{-1} A)^{-1} A^{\rm T} N^{-1} d,
\end{equation} 
where $N^{-1}$ represents the inverse of the time-domain noise covariance matrix. This can be calculated as 
\begin{equation}
N^{-1}=\mathcal{F}^{-1}[P(\omega)^{-1}] ,
\end{equation}
where $\mathcal{F}$$^{-1}$ represents the inverse Fourier-transformation and $P(\omega)$ is a matrix constructed from the auto- and cross-power spectra of the TODs, containing information about the detectors common-mode noise, in addition to the uncorrelated noise terms: 
\begin{equation}
P^{-1}(\omega)=[\alpha \langle c(\omega)^{\dagger}c(\omega) \rangle \alpha^t + \langle \tilde{n}^{\dagger}(\omega) \tilde{n}(\omega) \rangle ]^{-1}.
\end{equation}
The inverse of the pixel-pixel noise covariance matrix, $N^{-1}_{pp'}=(A^T N^{-1} A)^{-1} $ is not calculated explicitly. The mapmaker uses an iterative algorithm based on the conjugate gradient method with preconditioner to find the maximum likelihood solution for the map. 
Usually a few hundred iterations are needed to reach convergence.  
The computational time scales with the square of the number of bolometers
and also depends on the number of samples, $n_s$, in the TOD as $n_s {\rm log}(n_s)$.
Our observations consist of 34.5 hours of data for each bolometer sampled 
at a frequency of 18.6\,Hz. The 250\rmicron\ array has the largest number of 
bolometers (139) so the map created from this data has the longest 
processing time. Using eight 2.8\,GHz processors (Intel Xeon X5560 CPUs) the mapmaker needs about 
17 hours to reach convergence at 250\rmicron.

\begin{figure}[t!]
\centering
\vspace{-3.9mm}
\hspace{-4mm}
\includegraphics[width=.50\textwidth]{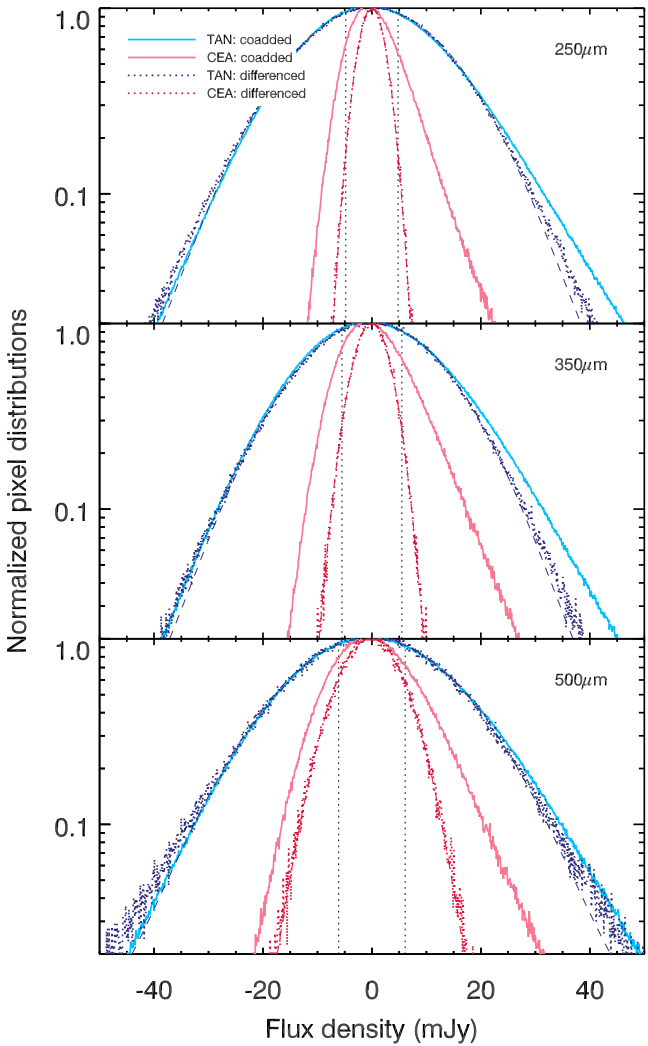}
\caption{
Pixel flux distributions of the coadded (i.e., the sky) and differenced jackknife maps, represented as solid and dotted lines, respectively, for both TAN (wider light/dark blue) and CEA (narrower red/pink) projections.   
Coadded maps include the entire data set in each band, while in the differenced maps the sky signal has been removed, leaving only instrumental noise.  
Coadded histograms are thus wider because they include confusion noise, represented by vertical dashed lines from \citet{nguyen2010}, as well as an excess compared with a  Gaussian at brighter flux densities from resolved sources.   
The full width at half maxima of the best-fit Gaussian to the difference maps --- shown as faint blue and red dashed lines for TAN and CEA , respectively --- represent average instrumental noise levels.    
Note that the TAN maps with their smaller pixels are dominated by instrumental noise, while the  bigger pixel CEA maps have approximately equal contributions from instrument and confusion noise.  
}
\label{fig:noise}
\end{figure}

\subsection{Noise properties}
\label{sec:np}

To examine the properties of the residual noise in our signal maps, we create \lq\lq jackknife\rq\rq\ difference maps, i.e., the timestream data are split into two halves and a separate map is made for each half, and the difference map is then made by multiplying one of the jackknifes by minus one and then averaging the two together. This process removes the astronomical signal but retains the noise, as the jackknife difference map contains the same instrumental noise properties as the coadded sky map.  
There are in principle several different ways to split the data in half, some more effective than others, but the shallow depth of the HerS observations in practice limits our options.  For example, since the field is only scanned once in each orthogonal direction,  we cannot split the TODs into two halves based on observation time, and splitting the datasets by orthogonal scan-direction results in maps that have strong residual correlated noise along the scan directions, due to lack of cross-linking. A third way to split the data is to divide up the detector focal planes, and only use every second bolometer to make our maps. Even though this method gives the best coverage, at the nominal pixel sizes the resulting maps are still quite sparse, especially at 500\rmicron\ where the sampling density is the lowest. This problem is not present in the larger pixel-size maps corresponding to the ACT mapping, and after correcting for the effect of the bigger pixel size we recover values similar to those in the more finely sampled maps.  

In Figure~\ref{fig:noise} we plot pixel-histograms of the coadded (or sky) and differenced jackknife maps --- in shades of blue for the standard (TAN) maps and red for the HerS-ACT (CEA) maps --- as solid and dotted lines, respectively.    
The coadded jackknife maps contain both instrument and confusion noise (the latter illustrated as vertical dotted lines), and are thus wider than the differenced jackknife maps.   
However, while the instrument noise is the dominant contribution in the TAN maps, the instrument noise in the HerS-ACT CEA maps is lower, by virtue of their pixels being 24.5, 12.7, $6.1\times $ larger (by area) at 250, 350, and 500\rmicron, respectively, such that  they have approximately equal contributions from instrument and confusion noise.

Instrumental noise levels are calculated by fitting a Gaussian to the pixel-histogram of the differenced jackknife maps for both the TAN and CEA cases. 
We find that the noise is extremely well described by the Gaussian fit (shown as thin dashed lines in the Figure~\ref{fig:noise}), 
deviating only at 500\rmicron\ by less than 2\%, and that the deviation is explained by the non-uniformity in  the samples per pixel arising from the sparseness of the array and the fact that we only cover each area with two scans.  
The resulting $1\sigma$ values in the TAN (CEA) maps are  11.9 (2.2), 11.4 (3.1), and 13.5 (5.4) $\rm mJy\, beam^{-1}$ at 250, 350, and 500\rmicron, respectively. 
Note that since the coverage of the HerS maps is not completely uniform (seen clearly in Figure~\ref{fig:hits}), the noise levels where more than two orthogonal scans overlap is lower.  
In the deeper regions of the TAN (CEA) maps, the noise levels are 
10.7 (2.1), 10.3 (2.8), and 12.3 (4.9) $\rm mJy\, beam^{-1}$, while in the shallower regions they are 
13.3 (2.5), 12.7 (3.4), and 14.9 (6.0) $\rm mJy\, beam^{-1}$ 
 at 250, 350, and 500\rmicron, respectively. 

{\sc sanepic} also creates an error map as an extension to the output products. This map gives an estimate of the variance of the noise in each pixel of the final map. Obtaining this error term correctly would require calculating the explicit pixel-pixel noise covariance matrix, but that operation is too computationally intensive and is never carried out during the iterative mapmaking. The error map {\sc sanepic} creates is a first-order estimate of this noise, computed by neglecting the off-diagonal terms in the inverse pixel-pixel noise covariance matrix, assuming that the final map only contains white noise. These determinations over-estimate the real residual noise values in the maps, but the error map can still be used to assign weights to each pixel in our final map. 

\begin{figure}[t!]
\centering
\vspace{2.6mm}
\hspace{-11mm}
\includegraphics[width=.530\textwidth]{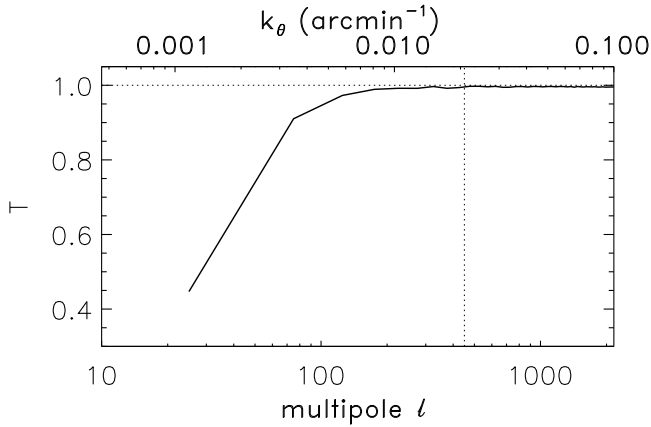}
\caption{Transfer function, $T$, of the {\sc sanepic} mapmaker at 500\rmicron, estimated with a Monte Carlo simulation as described in \S~\ref{sec:tf}.  
$T$ is found to be approximately unity down to $\ell \sim 200$ ($\sim 1\, \rm deg$), dropping to 0.5 at $\ell \sim 30$.  
The vertical dashed line represents the largest accessible scale, given the finite size of the survey, showing that effectively all scales in the map are reconstructed.  
Upper axis indicates $k_\theta \equiv \ell/(2\pi)$.  
}
\label{fig:tf}
\end{figure}
\begin{figure*}[t!]
\centering
\hspace{10mm}
\includegraphics[width=1.0\textwidth]{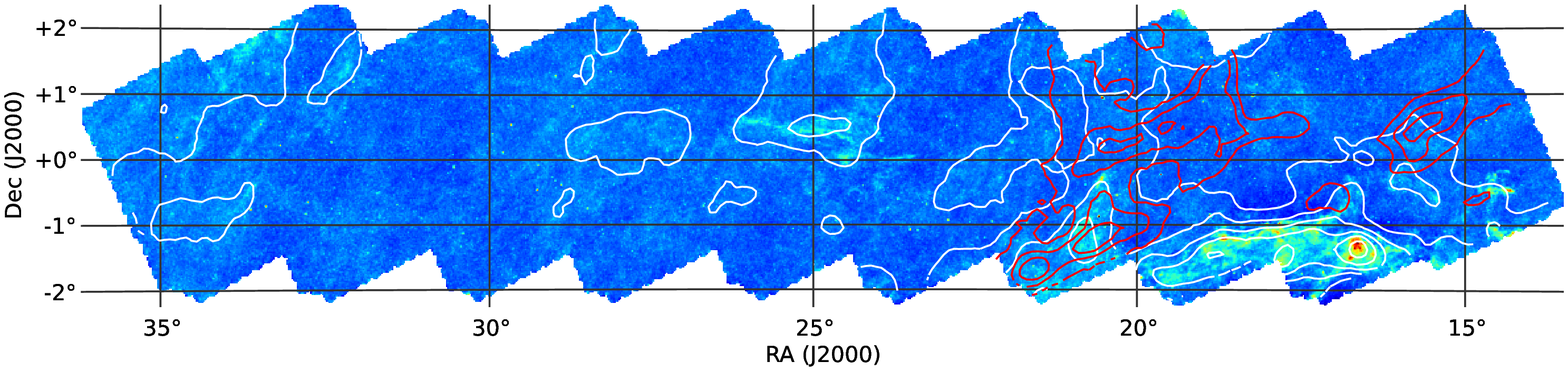}
 \caption{HerS 250\rmicron\ map, smoothed to 2\,arcmin, overlaid with contours representing the column density of local velocity clouds (white) and IVCs (red), as traced by \HI\  emission from GASS 21-cm data.  Note that no HVCs appear in this field.    White contours show $N_{\rm H}$ at 3.4, 4.2, 5.0, 5.8, 6.6, and $7.4\times 10^{21}\, \rm H\, cm^{-2}$, while red contours show $N_{\rm H}$ at 0.5, 0.8, and $1.1\times 10^{21}\, \rm H\, cm^{-2}$.   
 The color scale ranges linearly from -25 (blue) to 80\,mJy (red).    
The vast majority of the cirrus visible in HerS is attributable to the local velocity component.}
\label{fig:cirrus}
\end{figure*}
\subsection{Transfer Function}
\label{sec:tf}
We investigate how reliable our mapmaker is in reconstructing
large-scale structure on different angular scales. This assessment is made by
creating simulated pure-signal maps, which are then reprojected into
detector TODs and fed back into our mapmaker the same way as for the real
data. The ratio of the azimuthally-averaged Fourier transform of the
reconstructed map and the pure-signal input map gives us the mapmaker's
transfer function. In the ideal case the ratio should be unity at all
spatial scales. However, the mapmaker can introduce false signal to our
maps, or remove existing power, which would appear as a deviation from unity in
the transfer function. On the scales where the deviation from unity is not
too large, we can correct for these effects. We created 100 pure signal
maps with a power-law power spectrum  
resembling that of the cosmic infrared background without the
cirrus, and \lq\lq observed\rq\rq\ them with a Monte Carlo simulation at 500\rmicron,   
though we check that the transfer function is the same at all wavelengths with a small subset of simulated maps.  
Figure~\ref{fig:tf} shows the resulting transfer function. The mapmaker can
successfully reconstruct all large scales that are accessible in our maps. 
The simulated and reconstructed maps
were made with the same pixel size, so the pixel window function does not
have any effect here, and the transfer function remains unity on small scales.
The transfer function only starts to drop for $\ell \lsim 200$, corresponding to 
approximately half of the narrowest extent of our survey.
\subsection{Galactic Cirrus}
\label{sec:cirrus}
 Thermal emission by diffuse interstellar dust in our Galaxy --- the diffuse Galactic cirrus --- can be described by a modified blackbody proportional to $\nu^{\beta}B(\nu)$, where $B(\nu)$ is the Planck function and $\beta$ is the emissivity index, with temperatures ranging from 17 to 20\,K in the most diffuse regions \citep[e.g.,][]{boulanger1996,bracco2011}, to as low as 14\,K in dense regions where molecular hydrogen (\HII) can form \citep[e.g.,][]{netterfield2009,planck_xix},  peaking in emission between 150 and 200\rmicron.   

In diffuse regions, dust is well-traced by atomic hydrogen gas (\HI) which emits at 21\,cm and in the radio \citep[e.g.,][]{boulanger1996}.   
\HI\ data are available for HerS from the Parkes Galactic All-Sky Survey \citep[GASS;][]{mcclure2009,kalberla2010}, a 1.4\,GHz survey of Galactic atomic hydrogen emission, taken with the Parkes 64-m radio telescope, over $> 20{,}000\, \rm deg^2$.  
Publicly available data\footnote{The GASS second data release data server: \url{http://www.astro.uni-bonn.de/hisurvey/gass/}} 
are provided as velocity cubes, with effective angular resolution of $16\arcmin$, and velocity resolution of $1.0\, \rm km\, s^{-1}$.  
Following \citet[e.g.,][]{planck_xxiv}, we divide the cubes by velocity with respect to the local frame of reference into local ($|v_{\rm LSR}| < 30\,{\rm km\,s^{-1}}$) and intermediate velocity clouds (IVCs; $|v_{\rm LSR}| > 30\,{\rm km\,s^{-1}}$).   Note that no HVCs are visible in this field.  
We plot contours of the local and IVC components in Figure~\ref{fig:cirrus} (white and red contours, respectively), showing that the cirrus emission in HerS comes predominantly from the local velocity component.  

For column densities of roughly $N_{\rm H} \lsim 4 \times 10^{21}\, \rm cm^{-2}$, and on scales of an arcminute or less, this foreground is easily suppressed with a high-pass or matched filter \citep[e.g.,][]{chapin2011}.  High concentrations of \HI\ (and potentially \HII) --- such as that present in the bottom right corner of the HerS maps --- present a greater challenge; we describe our filtering method for point source identification in \S~\ref{sec:catalog}.   

\section{Catalog}
\label{sec:catalog}
We now present the first HerS band-merged catalog.  
We caution the user that because of the uneven coverage of the survey, the density of high signal-to-noise sources is higher in the deep stripes. 
Thus, though the catalog is suitable for applications such as cross-identification of bright objects, or cross-correlations given appropriate weights, etc., it should not be used for estimating statistically rigorous quantities such as source counts.  That does not mean that they cannot be measured, just that such operations are better done using the maps themselves, where the detailed properties of the extraction technique vs.\@ survey depth, etc., and the resulting catalog, can be properly simulated.  
\subsection{Catalogue production}
Point-source catalogs across the HerS field in the three SPIRE bands were produced using a three-step process: map filtering (to remove large-scale Galactic cirrus); source identification; and source extraction or photometry. We now describe the details of each of these steps in turn.

{\bf Filtering} of the HerS maps is done using a tapered high-pass filter that begins to remove power on scales larger than three times the beam FWHM at each SPIRE band. Specifically, we take the 2D Fourier transform of each map and attenuate spatial frequencies lower than $k=1/b,{\rm arcmin}^{-1}$ by a factor $(kb)^3$, where $b=3\times\mathrm{FWHM}$ in arcmin i.e., 
\begin{equation}
d_{\mathrm f}(x,y)= \begin{cases}
\mathcal{F}^{-1} k\ge1/b, & \hat{f}(l,m)\\
\mathcal{F}^{-1} k<1/b, & {\hat{f}(l,m)}{(kb)^3}\\
\end{cases},
\end{equation}
\noindent where $d_{\mathrm f}$ is the filtered map, $\hat{f}(l,m)$ is the Fourier transform of the observed map with frequencies $l$ and $m$ in the x and y directions, respectively, and $k=\sqrt{l^2+m^2}$.

\begin{figure*}[t!]
\centering
\includegraphics[width=0.925\textwidth]{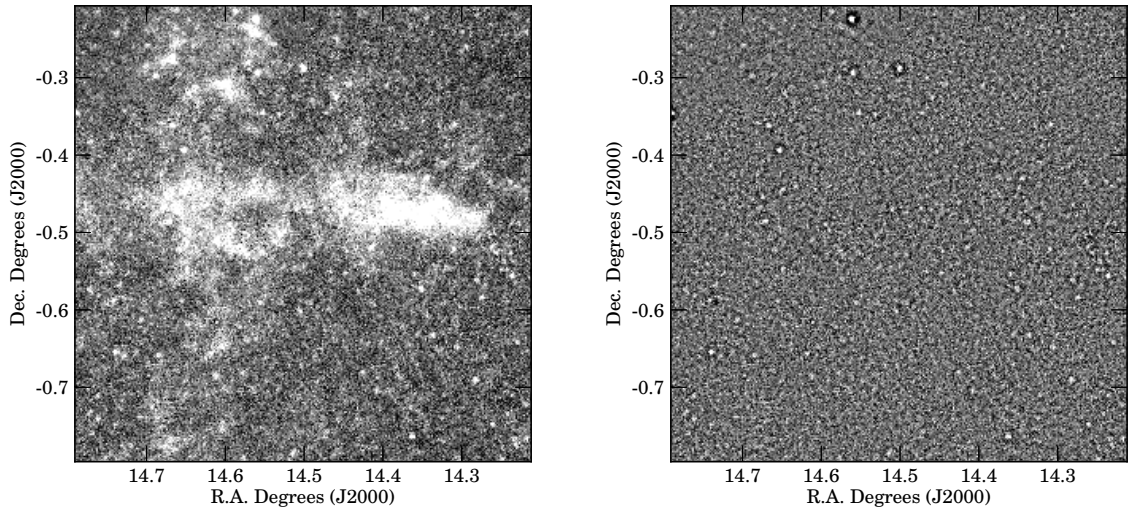}
\caption{$36\arcmin \times36\arcmin$ cutout from the HerS 250\rmicron\ map, in a region badly affected by Galactic cirrus. The cutout on the left is with no filtering, while the one on the right is with the tapered Fourier filtering described in the text. The effectiveness of the filter in removing large-scale features can be seen, but it also makes negative \lq\lq bowls\rq\rq\ around the brightest features, which must be accounted for by filtering the PSF when extracting sources.}
\label{fig:mapfiltering}
\end{figure*}

The minimum filtering scale of $(3\times\mathrm{FWHM})^{-1}$ was chosen to preserve as much of the source profile as possible while still suppressing any non-point like structure in the map. 
In Figure \ref{fig:mapfiltering} we illustrate the effectiveness of this filtering on a $36\arcmin \times 36\arcmin$ region of the HerS 250\rmicron\ image that is badly affected by cirrus contamination, with all power on scales larger than the beam efficiently suppressed.   
Consequently, negative \lq\lq bowls\rq\rq\ are visible around the brightest sources; next we describe how this is addressed when extracting point sources by filtering the point-spread function (PSF).  

{\bf Identification} of point sources in the filtered 250\rmicron\ image using the {\sc IDL} software package {\sc starfinder} \citep[][]{diolaiti2000}. Sources are assumed to be exclusively point-like in the SPIRE images, with a  PSF described by a circular 2D Gaussian with FWHM of 18.15, 25.15 and 36.3\,arcsec for 250, 350, and 500\rmicron, respectively. To account for the effect of our Fourier filtering (i.e., the \lq\lq bowls\rq\rq) the PSF is filtered in the same way as the map and this filtered PSF is used in the subsequent source detection and extraction steps. 
While {\sc starfinder} can operate in an ``iterative'' mode, detecting and removing sources at decreasing signal-to-noise ratio (SNR) thresholds, so as to allow the identification of faint sources in crowded regions, here we use a single pass of {\sc starfinder} requiring peak $\rm SNR > 3$ and $\rho_{\rm PSF}$, the correlation coefficient\footnote{$\rho_{\rm PSF}= \frac{\sum_i^N (d_i-\bar{d}))(P_i-\bar{P})}{N\sigma_d\sigma_P}$, where $d$ are the map pixel values and $P$ is the PSF}, to be greater than 0.5. In this setup {\sc starfinder} can be considered to be a simple peak finder; pixels in the map with $\rm SNR > 3$ are identified, collated into independent peaks, and then cross-correlated with the known PSF to confirm they are truly sources and not simply noise. Note that the uneven coverage of the maps, and subsequent deeper stripes (Figure~\ref{fig:hits}) with lower noise properties (\ref{sec:np}),  leads to a higher density of $\rm SNR > 3$ sources in the deep regions.     

{\bf Source photometry} is performed using a modified version of the De-blended SPIRE Photometry ({\sc desphot}) algorithm \citep[henceforth R12; Wang et al., in prep.\@]{roseboom2010, roseboom2012} developed for use on SPIRE data from the HerMES project \citep{oliver2012}. The main advantage of this approach is that it deals with the source blending issue in a way more appropriate to SPIRE maps than {\sc starfinder}, and produces consistent, band-merged SPIRE catalogues by using the input sources at the highest resolution band (250\rmicron) as a prior for the other SPIRE wavelengths.

While a complete description of how {\sc desphot} works is given in the above-listed papers,  we briefly summarize the main points here. 
For source photometry, {\sc desphot} assumes that the map (or each map segment) can be described as the summation of the flux density from the $n$ known sources in the map, i.e.,
\begin{equation}
{\bf d}=\sum_{i=1}^n{\bf P} f_i+\delta, 
\label{eqn:mlflux}
\end{equation}
where {\bf d} is the image data, {\bf P} the PSF for source $i$, ${f_i}$ the flux density of source $i$, and $\delta$ an unknown noise term. As discussed in \citet{roseboom2010} a linear equation of this form will (as in \S~\ref{sec:mm}) have a maximum likelihood solution 
\begin{equation}
{\bf \hat{f}}=({\bf A}^{\rm T}{\bf N_d}^{-1}{\bf A})^{-1}\, {\bf A}^{\rm T}{\bf N_d}^{-1}{\bf d},
\label{eqn:mlflux}
\end{equation}
where {\bf A} is an $m$ pixel by $n$ source matrix that describes the PSF for each source in the map and ${\bf N_d}$ is the noise covariance matrix. The best non-negative solution for ${\bf \hat{f}}$ is found using  the {\sc lasso} algorithm, as described in R12. As it is not computationally feasible to solve for the full set of $\sim30{,}000$ sources simultaneously, the input list must be broken up into ``groups" of sources that have significant overlap. In R12 this is accomplished by identifying high SNR ``islands'' in the SPIRE maps, but the HerS images are simply too big for this to be a reasonable option. Thus we group the {\sc desphot} input list with a ``friends-of-friends algorithm'', specifically the {\sc spheregroup} routine available as part of the SDSS {\sc idlutils}, using a linking length of 3\,arcmin. Friends-of-friends clustering algorithms have been used extensively in astronomy, typically for the identification of halos in dark matter simulations \citep[e.g.,][]{davis1985}. The algorithm works simply to uniquely group sources which are separated by less than the linking length. Groups are collated by identifying common neighbors (``friends'') so that each source belongs uniquely to one group. 

Despite the relatively shallow nature of the HerS observations, confusion is still a significant contributor to the noise budget for point sources. This complicates the selection criteria for a useful source catalogue as the point source detection stage described above isolates sources with a $\rm SNR >3$, taking into account only the instrumental noise. For example, at 250\rmicron\ point sources in the shallow (deep) region have a mean instrumental noise, estimated via error propagation of the hits map, of 7.7 (6.3)\,mJy, while the total noise, estimated via the pixel distribution of the point-source convolved map, is 11.1 (10.2)\,mJy. Note that these noise figures differ from those presented in Section \ref{sec:np}, as here we are considering the noise not in a single map pixel, but integrated over a point source. To proceed we follow a similar approach as \citet{smith2012}: the confusion noise is assumed to be constant across the entire map and is estimated via $\sigma_{\mathrm{conf.}}^2=\sigma_{\mathrm{total}}^2-\langle\sigma_{\mathrm{inst.}}\rangle^2$, where $\sigma_{\mathrm{total}}^2$ is the variance of the point source convolved map, and $\langle\sigma_{\mathrm{inst.}}\rangle$ is the mean instrumental noise in the map. The total noise for each source $i$ is then taken to be $\sigma_i^2=\sigma_{\mathrm{conf.}}^2+\sigma_{\mathrm{inst.},i}^2$. Using this approach we get $\sigma_{\mathrm{conf.}}=8$\,mJy for each of the 250, 350, and 500\rmicron\ band. These values are slightly higher than those presented by \citet{nguyen2010} and \citet{smith2012}; this is likely due to the effect of the Fourier filtering.  Using this definition of the the total noise, $\sigma$, for sources in our catalogue we threshold the catalogue to only include sources with 
$S_{250}>3\sigma$. For sources in the shallow regions this limit translates to $S_{250}\geq31$\,mJy while for the deep regions it is $S_{250}\geq28$\,mJy

\begin{figure}[t!]
\centering
\vspace{-3mm}
\includegraphics[width=.50\textwidth]{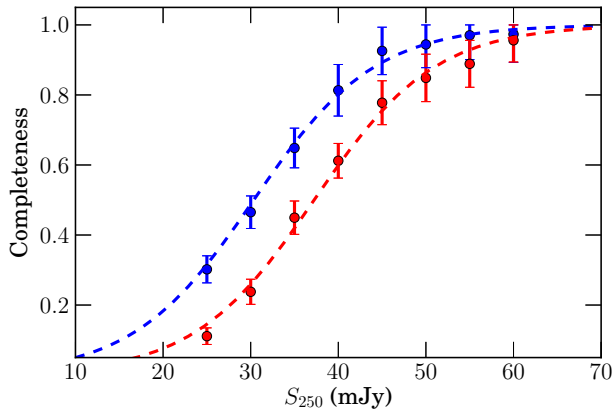}
\caption{HerS Catalog Completeness at 250\rmicron\ estimated by injecting grids of mock sources into the map. The blue points show the completeness in the deep region, while the red points are the shallow regions. The dashed lines are logistic fits to the data; $C=1/(1+\exp(0.145S_{250}+\beta))$, where $\beta=4.4$ for the deep regions, and $\beta=5.4$ for the shallow region.}
\label{fig:comp}
\end{figure}
\subsection{Completeness and Reliability}
\label{sec:completeness}
The completeness and reliability of the HerS catalogue is assessed using Monte Carlo techniques. The completeness is estimated by injecting grids of sources into the HerS maps and measuring the fraction that are detected (as $3\sigma$ sources) using the photometry pipeline. The input grids are matched to the output catalogue using a 6\,arcsec matching radius, which we estimate will produce spurious matches between unassociated input mock sources and real SPIRE sources at a rate of 0.5\%. As the HerS catalogue makes use of a 250\rmicron\ prior (i.e., we do not consider sources undetected at 250\rmicron) only the completeness at this wavelength is assessed. Figure~\ref{fig:comp} presents the completeness as a function of 250\rmicron\ flux density for the HerS catalogue in both the deep and shallow regions. It is reasonable to expect that the completeness, $C$, follows a logistic function, i.e., $C=1/(1+\exp\{\alpha S+\beta\})$. For both the deep and shallow completeness data we fit for the parameters $\alpha$ and $\beta$, finding $\alpha=0.145$ for both regions, while $\beta=4.4$ for the deep region and $\beta=5.4$ for the shallow region.

It is worth noting that this assessment of the completeness only considers the recoverability of sources at a given {\it true} flux density; at low SNR, the measured flux densities will be strongly affected by Eddington-type bias, i.e., $\langle S_{\rm obs}\rangle>\langle S_{\rm true}\rangle$. While the true impact of such flux boosting can only be assessed by taking into account the true distribution of flux densities \citep[i.e., the number counts;][]{coppin2006}, from our analysis we determine that  $S_{250}\sim 40\,$mJy is the faintest tested flux density at which the mean recovered flux density is equal to the injected value, i.e., $\langle S_{\rm obs}\rangle=\langle S_{\rm true}\rangle$.


 The reliability is estimated by taking jackknife realizations of the noise from deeper SPIRE imaging in the CDFS-SWIRE field. The HerMES observations of the CDFS-SWIRE field consist of eight scans of an 8 deg$^2$ region with SPIRE in fast scan mode. Thus we can produce four jackknife noise realizations at the depth of the HerS observations (2 scans) by producing maps from different pairs of scans in CDFS and subtracting away the eight scan maps. In order to assess the reliability of the HerS catalogue we run the pipeline on these noise-only maps. Across the four noise realizations (32\,deg$^2$) we detect 39 spurious sources at 3$\sigma$, giving a false positive rate of $1.2\pm0.2$\,deg$^{-2}$. Thus across the 79\,deg$^{2}$ of HerS we expect $96\pm16$ spurious sources.
\subsection{Details of the Published Catalog}
\label{sec:cat0}
Beginning with the catalog output by {\sc desphot}, we implement the following quality cuts: 
First we apply a 3$\sigma$ cut, where the completeness is estimated to be 50\% (from Figure~\ref{fig:comp})    
and false detection rate to be less than 1\%, as well as require reasonable residuals (i.e., $\chi^2<10$).    
Next, we identify obviously extended sources --- 24 in total --- where their extended nature results in them being broken up into multiple components by the filter, and remove them.  
This results in a catalogue with 32{,}815 sources at 250\rmicron, of which 13{,}300 and 3{,}276 have  similarly defined 3$\sigma$ detections at 350 and 500\rmicron, respectively.  

Sources fall in three distinct regions, identified with {\tt flag} in the catalog as either     
0) in the deep regions (16{,}626 sources); 1) in the wide regions (14,{083} sources); or 3) on the edges (2{,}106 sources). 
Wide regions are defined as those having the nominal coverage of two scans, while deep regions are those with three (and sometimes, but rarely, four) scans.   
Edges are the areas with only one scan of coverage.  Local counterparts of the extended sources are listed by name in the {\tt README} posted in the same directory.  
\section{Conclusion}
\label{sec:conclusion}
We present and make publicly available the first set of maps at 250, 350, and 500\rmicron, and catalog with $3.3\times 10^4$  sources detected at a significance of $\gsim 3\sigma$ (including confusion noise), from the \emph{Herschel} Stripe 82 Survey.  
Maps are made with the optimal mapmaker {\sc sanepic},  which we demonstrate recovers emission on all scales that are in principle accessible.  
The survey encompasses approximately half of the $150\, \rm deg^2$ of the deep SDSS Stripe in which Galactic foregrounds are subdominant at submillimeter wavelengths \citep[with HeLMS, described in][covering the other half]{oliver2012}.  Approximately $\sim 10\%$ of the HerS maps have significant foreground, with column densities $N_{\rm H} \gsim 4 \times 10^{21}\, \rm cm^{-2}$ and have been shown to be composed predominantly of local velocity clouds.   

The band-merged catalog is constructed, after filtering, with {\sc desphot} \citep{roseboom2010}, using 250\rmicron\ sources (extracted with {\sc starfinder}) as positional priors.   We include  sources with  
signal-to-noise greater than 3, 
whose completeness is estimated to be 50\% 
(Figure~\ref{fig:comp}),  
and false detection rate less than 1\%.  

HerS was designed with the intention of cross-correlating the maps with ancillary data --- whether maps or catalogs of galaxies or clusters --- to address a wide variety of questions.  It was initially proposed  to correlate with HETDEX Lyman $\alpha$ emitters (LAEs) at $1.8 < z < 3.5$ \citep[e.g.,][]{hill2008,adams2011} with the aim of measuring the contribution to the CIB from that redshift range and infer the star-formation rate density through this critical epoch.  Furthermore, combining that measurement  with stellar masses of LAEs estimated from the SHELA/SpIES catalogs, specific star-formation rates, and the relationship of star-formation to halo mass at higher-$z$ can be explored.   

Other exciting projects that we intend to pursue include: determining the correlation between HerS sources and clusters or cluster members, e.g., exploring the correlation of infrared emitting sources and clusters detected by ACT using the SZ effect  \citep[][]{hasselfield2013}; 
the lensing of the CMB by foreground structure traced by the CIB \citep{holder2013,planck_xviii, hanson2013}, and investigating the effect that the environment has on star-formation in sources identified as cluster members \citep{geach2012,rykoff2013}.   
SDSS/BOSS offers a wealth of galaxy and quasar  \citep[e.g.,][]{ross2009a, paris2012} populations for cross-correlation.

In addition to cross-correlations, single-object lensed or highly luminous high-redshift sources can be selected from the maps themselves.  By linearly combining the maps, high-redshift \lq\lq red peakers\rq\rq\ \citep[e.g., with $S_{250} < S_{350} < S_{500}$ at $z\gsim 3$; ][]{dowell2014,riechers2013} are identifiable.  High-redshift groups and clusters can be selected as red overdensities (e.g., the \emph{Planck} clumps; Clements et al., submitted), 
which alternatively can be used to clean the CIB from CMB maps to probe the damping tail of the CMB power spectrum \citep[e.g.,][]{hajian2012,keisler2011,reichardt2012,sievers2013}.

Studies focused on our Galaxy are possible as well.  The large-scale fidelity of our maps, as demonstrated by the transfer function shown in Figure~\ref{fig:tf}, allows large-scale properties of cirrus and dense molecular regions to be fully reconstructed, while our relatively small beam means that finer structures can be separated out. And by correlating dust emission in the infrared with measurements from optical fibers pointed at  \lq\lq blank sky\rq\rq, we can recover the optical spectrum of the diffuse Galactic light to constrain the size distribution of Galactic dust \citep[e.g.][]{brandt2012}.

Finally, future cosmological surveys such as the Dark Energy Survey (DES), Hyper Suprime-Cam (HSC), and the Large Synoptic Survey Telescope (LSST) 
will further enrich the density and variety of sources with which these submillimeter data can be cross-correlated, making this survey an integral component of an important Legacy field.

\begin{acknowledgments}
The authors warmly thank Duncan Hanson, Brandon Hensley, Edward Chapin, and Lyman Page for their input and participation.   We also thank the anonymous referee, whose comments have greatly improved this paper.  
\input{spire_acknowledgement.tex}
\end{acknowledgments}
\bibliographystyle{apj}
\bibliography{refs.bib}


\end{document}

%% file: spire_acknowledgement.tex
SPIRE has been developed by a consortium of institutes led
by Cardiff Univ. (UK) and including: Univ. Lethbridge (Canada);
NAOC (China); CEA, LAM (France); IFSI, Univ. Padua (Italy);
IAC (Spain); Stockholm Observatory (Sweden); Imperial College
London, RAL, UCL-MSSL, UKATC, Univ. Sussex (UK); and Caltech,
JPL, NHSC, Univ. Colorado (USA). This development has been
supported by national funding agencies: CSA (Canada); NAOC
(China); CEA, CNES, CNRS (France); ASI (Italy); MCINN (Spain);
SNSB (Sweden); STFC, UKSA (UK); and NASA (USA).